\documentclass{article}
\usepackage{spconf}
\usepackage[per-mode=symbol,per-symbol=p]{siunitx}
\usepackage{cite}
\usepackage{adjustbox}
\usepackage[ruled,vlined]{algorithm2e}
\usepackage{amsmath,amssymb,amsfonts}
\usepackage{algorithmic}
\usepackage{graphicx}
\usepackage{textcomp}
\usepackage[colorlinks=false, pdfborder={0 0 0}, breaklinks]{hyperref}
\usepackage{xcolor}
\usepackage{balance}
\usepackage{enumitem}
\setlist{nolistsep,leftmargin=*}
\usepackage{setspace}
\usepackage{wrapfig}
\usepackage{listings}
\usepackage{color}
\usepackage{caption}
\usepackage{subcaption}
\usepackage{todonotes}
\usepackage[longtable]{multirow}
\usepackage{makecell}

\def\BibTeX{{\rm B\kern-.05em{\sc i\kern-.025em b}\kern-.08em
    T\kern-.1667em\lower.7ex\hbox{E}\kern-.125emX}}

\usepackage{amsmath}
\DeclareMathOperator*{\argmax}{arg\,max}
\DeclareMathOperator*{\argmin}{arg\,min}

\usepackage{tikz}
\usetikzlibrary{shapes.geometric, arrows,shadows}
\usetikzlibrary{fit}
\usetikzlibrary{calc}
\tikzstyle{process_g} = [rectangle, minimum width=2.3cm, minimum height=0.75cm, text centered, text width=2.3cm, draw=black, fill=gray!40]

\tikzstyle{process} = [rectangle, minimum width=2.3cm, minimum height=0.75cm, text centered, text width=2.3cm, draw=black, fill=gray!10]
\tikzstyle{arrow} = [thick,->,>=stealth]

\newcommand*\circled[1]{\tikz[baseline=(char.base)]{%
            \node[shape=circle,fill=white!20,draw,inner sep=0.5pt] (char) {#1};}}
            
\usepackage{ctable}

\newcommand{\ie}{\emph{i.e.}, }

\newcommand{\HEVC}{\emph{High Efficiency Video Coding }}
\newcommand{\VVC}{\emph{Versatile Video Coding }}

\newcommand{\scheme}{\texttt{VEXUS}\xspace}

\newcommand{\EY}{$E_{\text{Y}}$}
\newcommand{\EU}{$E_{\text{U}}$}
\newcommand{\EV}{$E_{\text{V}}$}
\newcommand{\LY}{$L_{\text{Y}}$}
\newcommand{\LU}{$L_{\text{U}}$}
\newcommand{\LV}{$L_{\text{V}}$}
\newcommand{\h}{$h$}

\newcommand{\BDRP}{$BDR_{\text{P}}$}

\newcommand{\BDRX}{$BDR_{\text{X}}$}

\newcommand{\vig}[1]{\textcolor{black}{#1}}

\begin{document}
\title{Convex-hull Estimation using XPSNR for Versatile Video Coding}

\name{Vignesh V Menon \quad Christian R. Helmrich\quad Adam Wieckowski \quad Benjamin Bross \quad Detlev Marpe}
\address{Video Communication and Applications Department, Fraunhofer HHI, Berlin, Germany}

\maketitle

\begin{abstract}
As adaptive streaming becomes crucial for delivering high-quality video content across diverse network conditions, accurate metrics to assess perceptual quality are \vig{essential}. This paper explores \vig{using} the eXtended Peak Signal-to-Noise Ratio (XPSNR) metric as an alternative to the popular Video Multimethod Assessment Fusion (VMAF) metric for determining optimized bitrate-resolution pairs in the context of \VVC (VVC). Our study is rooted in the observation that XPSNR shows a superior correlation with subjective quality scores for VVC-coded Ultra-High Definition (UHD) content compared to VMAF. We predict the average XPSNR of VVC-coded bitstreams using spatiotemporal complexity features of the video and the target encoding configuration and then determine the convex-hull online. On average, the proposed con\underline{vex}-hull \underline{u}sing XP\underline{S}NR (\scheme) achieves an overall quality improvement of \SI{5.84}{\decibel} PSNR and \SI{0.62}{\decibel} XPSNR while maintaining the same bitrate, compared to the default UHD encoding using the VVenC encoder, accompanied by an encoding time reduction of \SI{44.43}{\percent} and a decoding time reduction of \SI{65.46}{\percent}. This shift towards XPSNR as a guiding metric shall enhance the effectiveness of adaptive streaming algorithms, ensuring an optimal balance between bitrate efficiency and perceptual fidelity with advanced video coding standards.
\end{abstract}


\section{Introduction}
Modern video standards come with ever-increasing complexity. The newest, \VVC~(VVC)~\cite{overview_vvc}, \vig{was expected} to be up to ten times more complex to encode than its predecessor, \HEVC~(HEVC)~\cite{overview_hevc}. Practical implementations, like the open source Versatile Video Encoder (VVenC)~\cite{vvenc_ref}, can efficiently deal with scaling the compression efficiency versus the runtime~\cite{preset_ref}  using various presets offering a trade-off between runtime and compression efficiency. The convex hull is where the encoding point achieves Pareto efficiency. Online convex-hull estimation methods emerge as a critical necessity in this landscape, providing a dynamic and adaptive means to optimize bitrate and resolution selections. By dynamically adjusting the bitrate-resolution pairs in response to the video content complexity and coding algorithms, these methods play a pivotal role in achieving an optimal trade-off between computational efficiency and visual fidelity in the face of the increased intricacies associated with advanced codecs like VVC~\cite{vvc_convexhull_ref,mcbe_ref}. 

\begin{figure}[t]
\centering
\begin{subfigure}{0.230\textwidth}
    \centering
    \includegraphics[width=\textwidth]{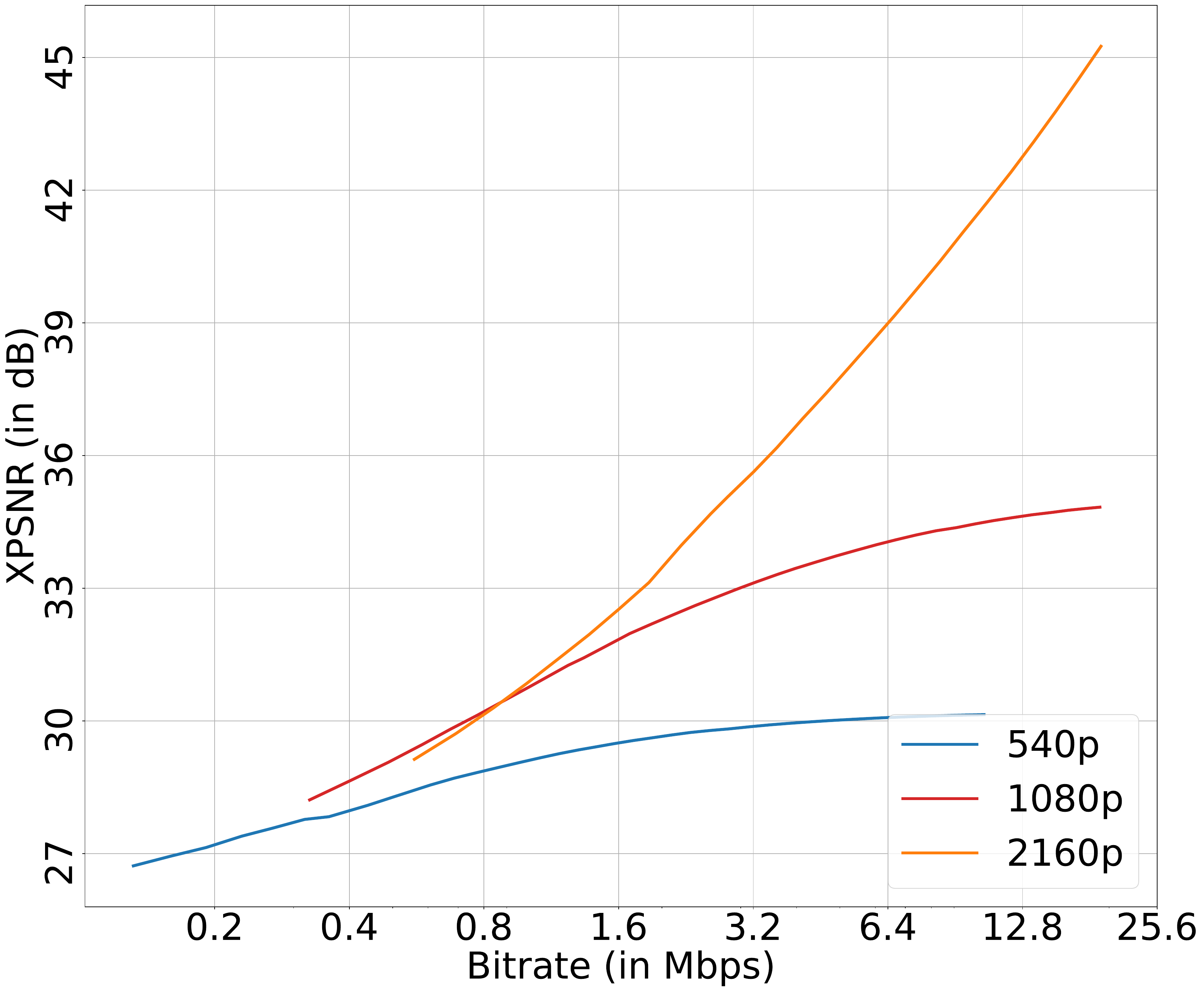}
    \includegraphics[width=\textwidth]{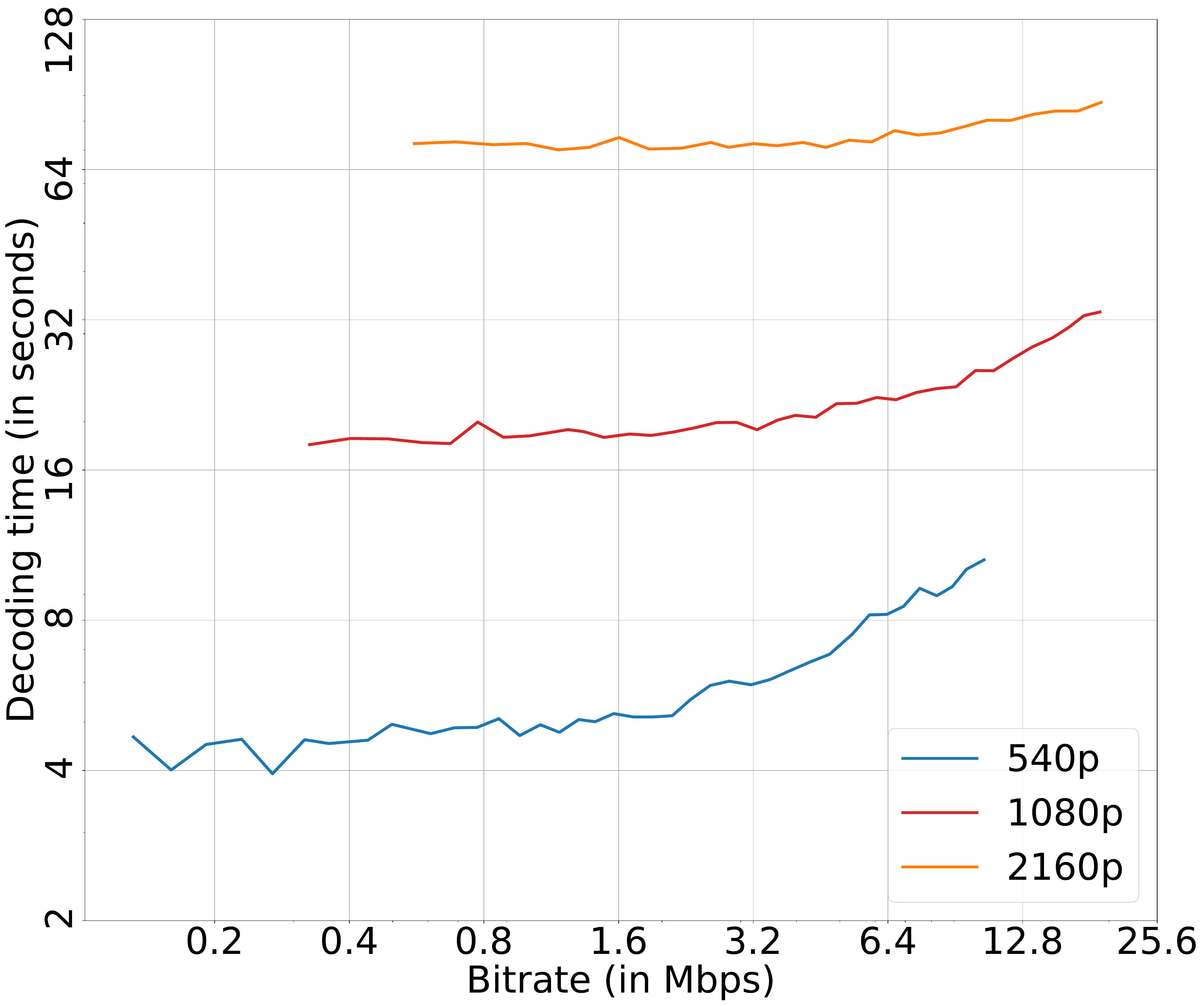}    
    \caption{\textit{0024}}    
    \label{fig:0024_intro}
\end{subfigure}
\begin{subfigure}{0.230\textwidth}
    \centering
    \includegraphics[width=\textwidth]{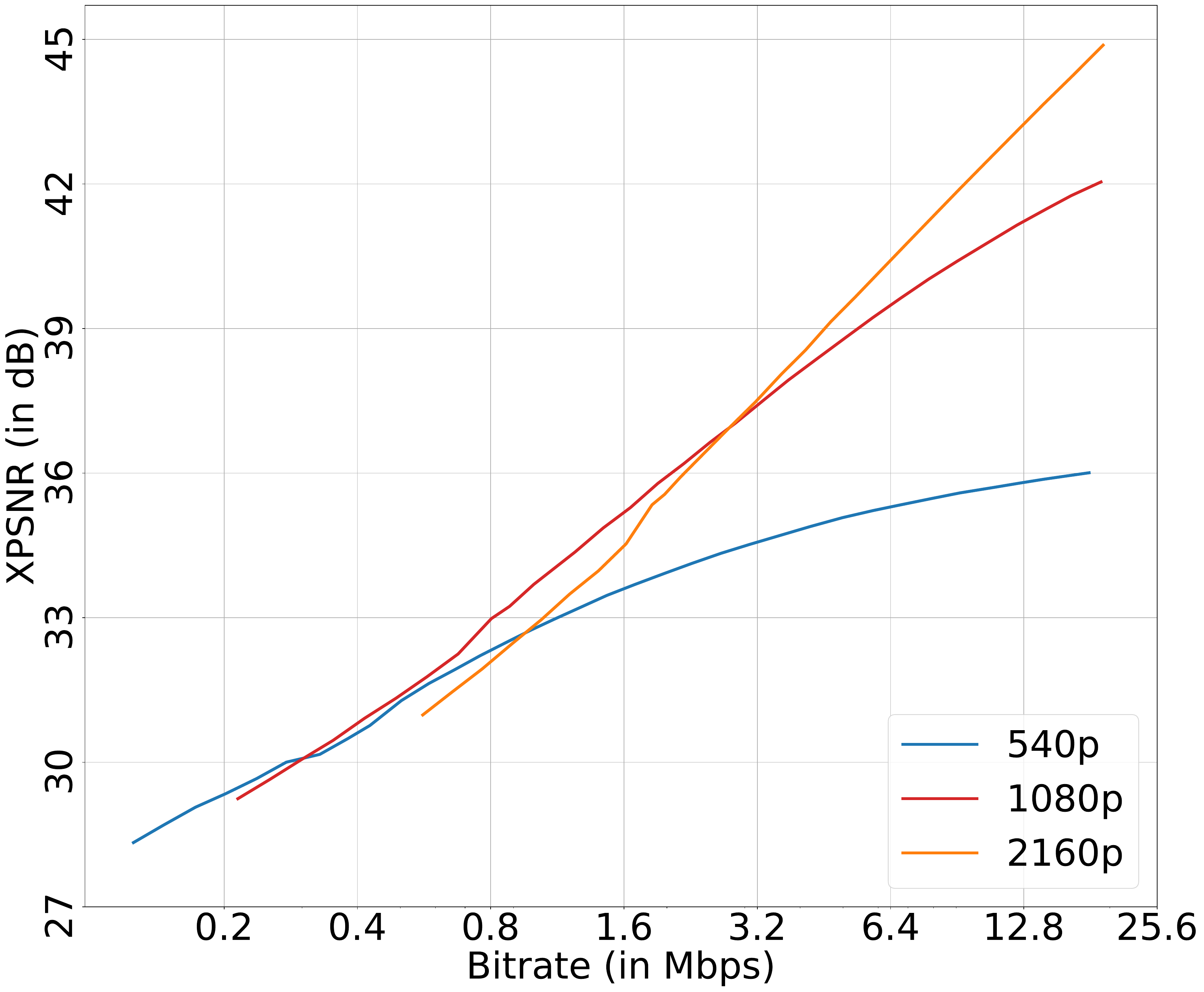}
    \includegraphics[width=\textwidth]{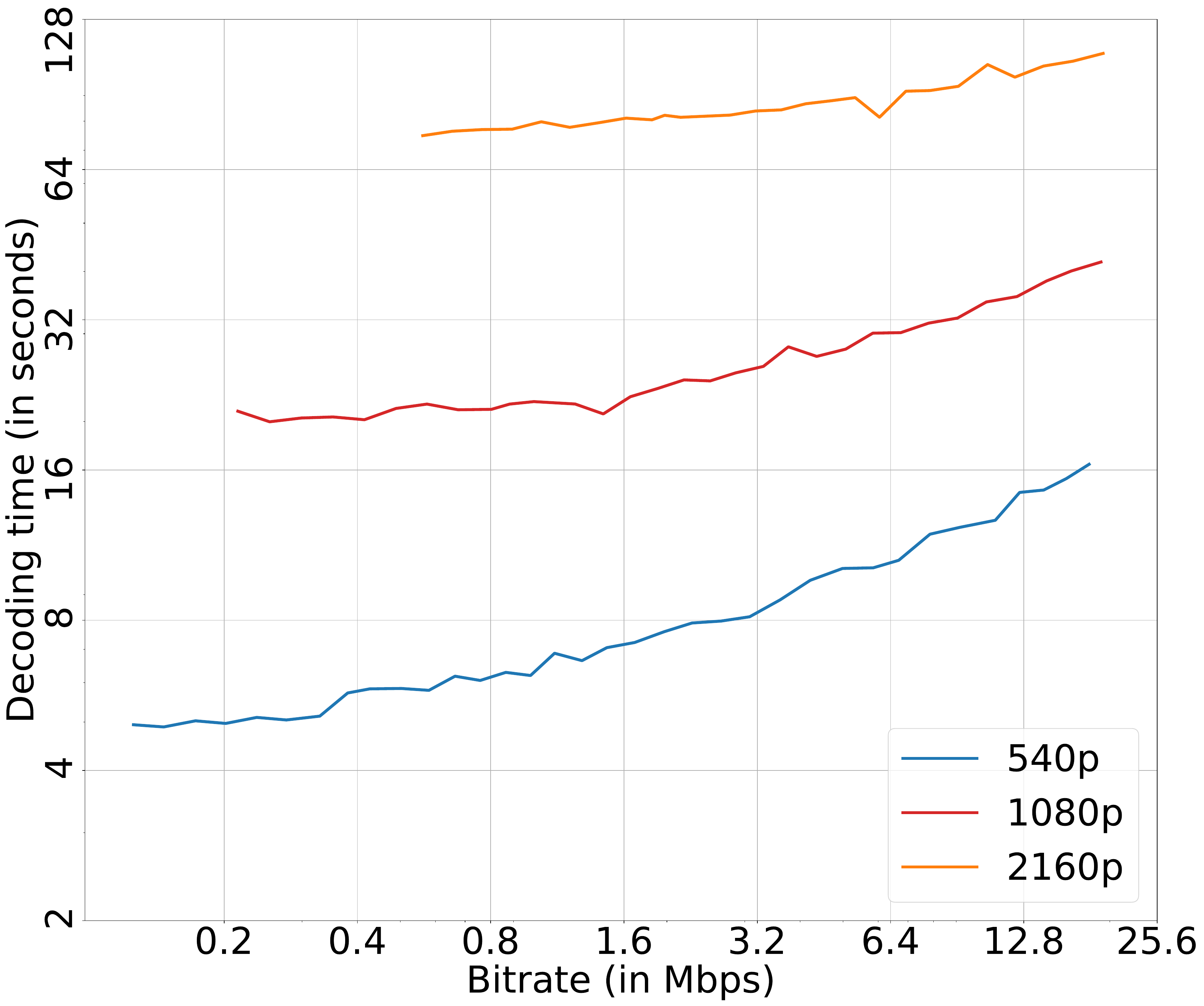}    
    \caption{\textit{0047}}    
    \label{fig:0047_intro}
\end{subfigure}
\vspace{-0.8em}
\caption{Rate-distortion (RD) and rate-decoding time curves of representative sequences of Inter-4K dataset~\cite{inter4k_ref} encoded at 540p, 1080p, and 2160p resolutions using VVenC at \textit{faster} preset, and decoded using VTM decoder.}
\vspace{-1.5em}
\label{fig:intro_convexhull}
\end{figure}

\begin{figure*}[t]
\centering
\includegraphics[width=0.98\linewidth]{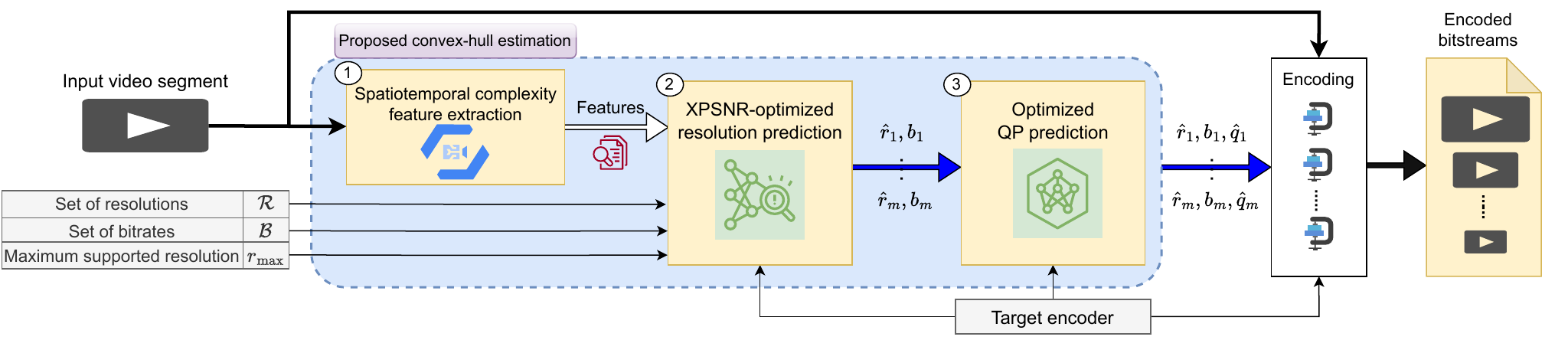}
\vspace{-1.045em}
\caption{Convex-hull estimation and encoding architecture using \scheme.}
\vspace{-1.05em}
\label{fig:contribution}
\end{figure*}

In content-adaptive encoding, especially for adaptive streaming applications, pursuing optimal bitrate-resolution pairs is crucial for balancing efficient video delivery and perceptual quality~\cite{netflix_paper}. The Pareto-front (referred to as \emph{convex hull} in the adaptive streaming literature) estimation methods used to determine the optimized bitrate-resolution pairs for individual video titles have been studied intensively in recent literature, motivated by the observation that specific encoding resolutions yield better quality than others in a certain bitrate range, as shown in Fig.~\ref{fig:intro_convexhull}. By considering the intricacies of content complexity, these methods ensure that encoding is precisely tuned to the requirements of each video, minimizing the bitrate and maximizing visual fidelity. Moreover, by selecting the optimized encoding resolution for each bitrate, we reduce the computational workload during decoding, as lower-resolution videos require less processing power and memory than their higher-resolution counterparts~\cite{visor_ref,dec_comp_vvc_ref}, as shown in Fig.~\ref{fig:intro_convexhull}. 
One of the convex-hull estimation methods is \mbox{\emph{Bruteforce}} approach, where, for a set of $\tilde{r}$ resolutions and $\tilde{q}$ quantization parameters (QP), video is compressed $\tilde{r}\times \tilde{q}$ times, and then the best resolution-QP pair is selected for each target bitrate~\cite{netflix_paper,chen_pte_ref}. Notably, the convex-hull construction is repeated for every scene. Some other techniques involve spatiotemporal complexity-adaptive encoding resolution selection, employing machine learning models to predict optimal encoding configurations~\cite{gnostic, faust_ref, res_pred_ref1,jtps_ref}. As the demand for high-quality and low-bitrate encoding intensifies, these cutting-edge convex-hull estimation methods stand at the forefront, promising more efficient and quality-driven video streaming experiences.

Traditional convex-hull estimation methods use Video Multimethod Assessment Fusion (VMAF)~\cite{VMAF} as the perceptual quality metric~\cite{netflix_paper,faust_ref,nasiri_pte_ref,jtps_ref}. However, as observed in~\cite{xpsnr_vs_vmaf}, Peak Signal to Noise Ratio (PSNR)~\cite{psnr_ref} and VMAF~\cite{VMAF} measures fail to model the subjective quality of VVC-coded bitstreams accurately. Moreover, it was observed that Structural Similarity Index (SSIM)~\cite{ssim_ref}, Multi-Scale Structural Similarity Index (MS-SSIM), and eXtended Peak Signal-to-Noise Ratio (XPSNR)~\cite{xpsnr_ref} can predict the subjective codec ranking reported in~\cite{wien_xpsnr_vs_vmaf} with acceptable accuracy~\cite{itu_xpsnr_vs_vmaf}. 
Hence, XPSNR appears to be a superior perceptual quality metric to VMAF when applied to VVC-coded ultra high definition (UHD) bitstreams~\cite{xpsnr_vs_vmaf}. XPSNR goes beyond traditional PSNR metrics by incorporating additional perceptual considerations, providing a more nuanced and human-like video quality assessment. 
XPSNR excels in aligning with the complexities introduced by VVC encoding, making it a more reliable and comprehensive metric for assessing the true perceptual quality of VVC-coded bitstreams compared to the more conventional VMAF. However, despite the efficiency gained by the faster calculation of XPSNR compared to VMAF~\cite{xpsnr_ref}, the computational demands of encoding videos at all supported resolutions and QPs, coupled with the subsequent XPSNR calculations, pose significant time and energy challenges. 

Encoding and evaluating an exhaustive set of configurations becomes increasingly burdensome as the number of supported resolutions grows. To mitigate this computational bottleneck, we propose a method, \scheme, which predicts optimized convex-hulls (using XPSNR as the quality metric) for VVC encoding without requiring exhaustive encoding and quality evaluation. It includes introducing XPSNR and QP prediction models that help \vig{estimate} optimized bitrate-resolution-QP triples for a given video scene.

\section{\scheme architecture}
\scheme streamlines the VVC encoding process to achieve optimal bitrate-quality trade-offs by introducing a comprehensive convex-hull estimation scheme with three distinct phases \vig{as shown in Fig.~\ref{fig:contribution}}:
\begin{enumerate}[topsep=0pt,leftmargin=*,label=\protect\circled{\arabic*}]
  \item Spatiotemporal complexity feature extraction;
  \item XPSNR-optimized resolution prediction;
  \item Optimized QP prediction.
\end{enumerate}
\scheme begins with spatiotemporal feature extraction, where intricate details and motion dynamics are analyzed to capture the inherent complexity of the input video scene. Subsequently, in the optimized resolution estimation phase, \scheme leverages the extracted features and a set of supported resolutions ($\mathcal{R}$) and a set of target bitrates ($\mathcal{B}$) to determine the most suitable encoding resolution. Finally, in the optimized QP estimation phase, \scheme refines its encoding strategy by incorporating target bitrates and previously estimated resolutions to predict the optimized QP values for each bitrate-resolution pair. Through this iterative approach, \scheme outputs bitrate-resolution-QP triples tailored to each target bitrate, ensuring efficient bitrate allocation and preserving visual quality across diverse resolutions in VVC encoding.

\subsection{Spatiotemporal complexity feature extraction}
\label{sec:features}
We use seven DCT-energy-based features: the average texture energy (\EY), the average gradient of the luma texture energy (\h), the average luma brightness (\LY), Average chroma texture energy of U and V channels (\EU~and \EV), and the average chroma \vig{intensity} of U and V channels (\LU~and \LV)~\cite{vca_ref} as the content complexity features of video scenes.

\subsection{XPSNR-optimized resolution prediction}
\label{sec:res_pred}
The objective of selecting the optimized resolution based on bitrate and video complexity features is decomposed into two parts:

\paragraph*{Modeling:} 
The core of the XPSNR measure is a psycho-visually inspired, simplified spatiotemporal sensitivity model specified locally for \vig{each disjoint block $B_{k}$ at index $k$}, of each input picture $P$ of bit depth $d$, and size $W\times H$ pixels, as~\cite{xpsnr_ref}:
\begin{align}
\resizebox{0.72\hsize}{!}{ $w_{k} = \sqrt{\frac{\hat{a}_{P}}{\hat{a}_{k}}},\hspace{1em} \text{with}\hspace{1em} \hat{a}_{P} = 2^{2d-9}\cdot \sqrt{\frac{3840\cdot 2160}{W\cdot H}}$ } \\
\resizebox{0.90\hsize}{!}{ \label{eq:a_k} $\hat{a}_{k} = \max\Big({\hat{a}^2}_{\text{min}}; \big(\frac{1}{4N^{2}} \sum_{[x,y]\in B_{k}} \mid h_{s}[x,y]\mid + 2 \mid h_{t}[x,y]\mid\big)^{2} \Big)$ }
\end{align}
\vig{where $w_{k}$ is the visual sensitivity weight associated with the $N\times N$ sized $B_{k} \in P$ and calculated from the visual activity measure of the block $\hat{a}_{k}$ and an average overall activity $\hat{a}_{P}$. $N$ is calculated as:}
\begin{align}
\label{eq:N}
\resizebox{0.47\hsize}{!}{   \vig{$N = \text{round}\Big(128 \cdot \sqrt{\frac{W\cdot H}{3840\cdot 2160}}\Big) $}.}
\end{align}

The definitions of $a_{\text{min}}$, $h_{\text{s}}$, and $h_{\text{t}}$ are provided in~\cite{itu_xpsnr_vs_vmaf} and omitted here for brevity, and $N^{2}$ is the number of picture samples in each block. A low computational complexity is reached because the spatial
high-pass operator $h_{\text{s}}$, and temporal high-pass operator $h_{\text{t}}$ use very simple fixed-point operations, and the square root in the calculation of $w_{k}$ cancels the squaring operations in Eq.~\ref{eq:a_k}.

Towards this realization, XPSNR of a video scene encoded at resolution $r$ and bitrate $b$, \ie $x_{(r,b)}$ is modeled as a function of video complexity features \{\EY, \h, \LY, \EU, \LU, \EV, \LV \}, $b$, and normalized resolution height $r`=\frac{r}{2160}$:
\begin{align}
\label{eq:v_pred}
\resizebox{0.67\hsize}{!}{   $ x_{\left(r,b\right)} = f_{\text{X}}\left(E_{\text{Y}}, h, L_{\text{Y}}, E_{\text{U}}, L_{\text{U}}, E_{\text{V}}, L_{\text{V}}, b, r`\right)$.}
\end{align}

\LY, \LU, and \LV~features consider variations in luminance and chrominance within localized regions. \EY, \EU, and \EV~features account for variations in texture across different frame regions, providing insights into how well the compression or reconstruction method preserves textural details. \h~introduces a temporal dimension, where dynamic changes in texture over time influencing perceived quality are captured~\cite{vqtif_ref,jtps_ref}.

\paragraph*{Resolution optimization:} 
In the context of resolution optimization, the goal is to enhance the XPSNR of encoded video scenes by predicting the optimized encoding resolution. This optimization process aims to select the resolution that maximizes the predicted XPSNR, thereby improving the visual quality of the encoded content. The optimization function is:
\begin{align}
   \hat{r} = \argmax_{r \in \mathcal{R}; r < r_{\text{max}}} \hat{x}_{(r,b)},  
\end{align}
where $\hat{x}_{(r,b)}$ is the predicted XPSNR of the video scene encoded at resolution $r$ and bitrate $b$. By iteratively evaluating the XPSNR prediction for different bitrate-resolution combinations, the optimization function identifies the resolution that yields the highest predicted XPSNR, ensuring an optimal balance between bitrate utilization and visual fidelity in the encoded video scenes.

\subsection{Optimized QP prediction}
\label{sec:qp_pred}
For applications such as streaming, avoiding exceeding the maximum bitrates specified in the HLS/DASH manifests~\cite{mpeg_dash_ref} during the encoding process is essential. Failure to adhere to these limits can lead to buffer overflows or underflows in video players. Therefore, optimized QP is predicted for a given target bitrate for capped variable bitrate encoding.

\paragraph*{Modeling:}
The optimized QP is modeled as a function of spatiotemporal features, target bitrate $b$, and normalized resolution height $r`$ as:
\begin{align}
\label{eq:q_pred}
\resizebox{0.67\hsize}{!}{   $ q_{\left(r,b\right)} = f_{\text{Q}}\left(E_{\text{Y}}, h, L_{\text{Y}}, E_{\text{U}}, L_{\text{U}}, E_{\text{V}}, L_{\text{V}}, b, r`\right)$.}
\end{align}
By considering spatiotemporal features, the encoding can adaptively adjust the QP parameter to efficiently allocate bits to different scenes within the video~\cite{allintra_rc_ref}. This ensures that the encoding gives more bits to preserve essential details in complex scenes while allocating fewer bits to simpler regions, optimizing the overall bitrate distribution. Target bitrate and resolution features serve as quality constraints and objectives for encoding. By incorporating these features into the prediction model, the encoding process can dynamically adjust the QP parameter to achieve the desired bitrate while maintaining the specified resolution. 

\paragraph*{Optimization:}
The optimization function is:
\begin{align}
   \hat{q}_{(r,b)} &= \argmin_{q \in [q_{\text{min}},q_{\text{max}}]} \mid \hat{b}_{(r,q)}- b \mid,  
\end{align}
where $\hat{q}_{(r,b)}$ is the predicted QP of the scene encoded at resolution $r$ and bitrate $b$. The optimization function aims to predict the QP, minimizing the discrepancy between the predicted and target bitrate for a given resolution. This function seeks to find the optimized QP value within the specified range $[q_{\text{min}}, q_{\text{max}}]$ that best aligns the predicted bitrate $\hat{b}_{r,q}$  with the target bitrate $b$. The chosen QP value is expected to result in an encoded scene with a bitrate that closely matches the specified target bitrate, thereby optimizing the encoding process for bitrate control.

\subsection{Encoding}
\vig{The encoding uses the predicted bitrate-resolution-QP configurations for a given input scene. In VVenC~\cite{vvenc_ref}, the QP is specified using the \texttt{qp} option. The \texttt{maxrate} (easy mode) or \texttt{MaxBitrate} (expert mode) option is used to specify the upper bound of bitrate variability. The lower bound of the bitrate variability is indirectly achieved through the XPSNR-based perceptual QP adaptation in combination with the \texttt{qp} option~\cite{vvenc_rc_ref}. This means that while the maximum bitrate is explicitly set, the encoder adjusts the QP values dynamically based on the perceptual quality model to avoid exceeding the lower bitrate threshold. This adaptive adjustment ensures that the bitrate remains within the acceptable range without requiring explicit specification.}

\section{Experimental design}
\label{sec:exp_eval}

\subsection{Dataset preparation}
\begin{figure}[t]
\centering
\includegraphics[clip,width=0.995\linewidth]{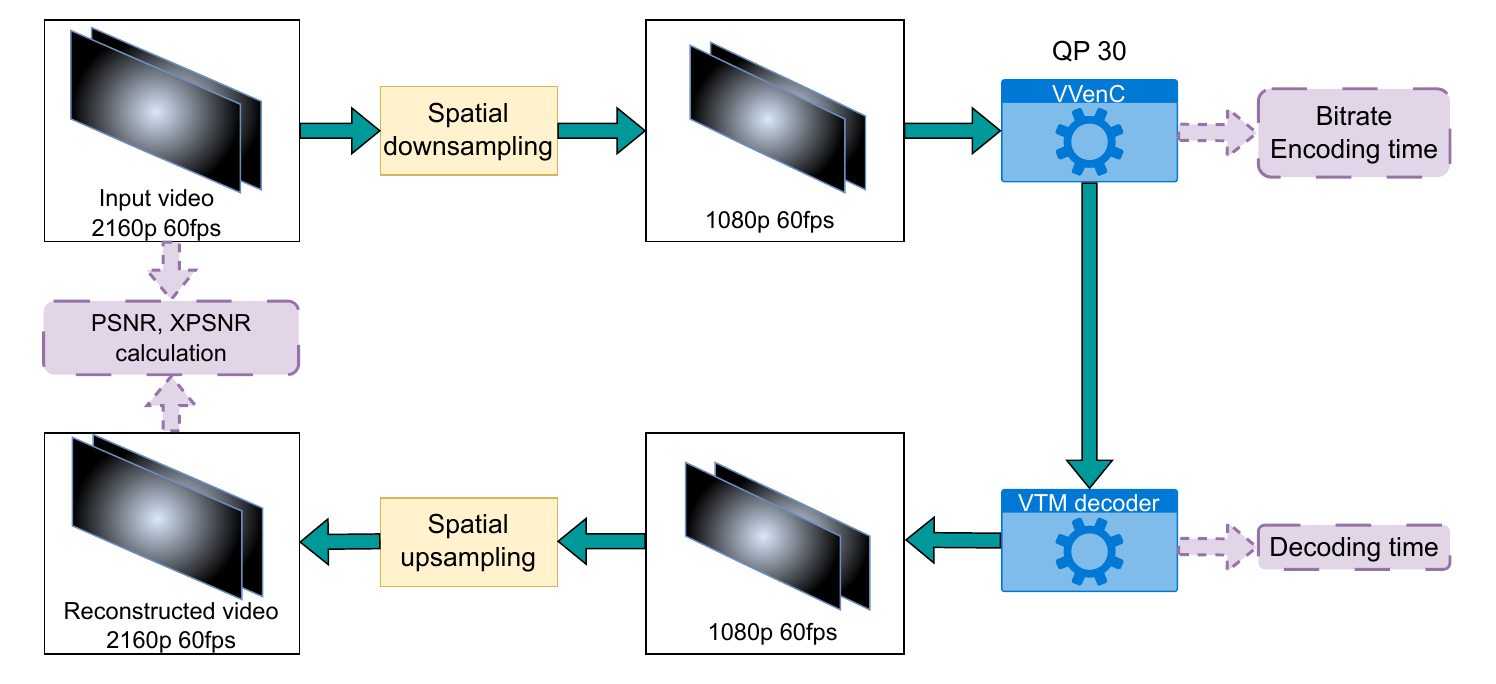}
\vspace{-0.9em}
\caption{Calculation of the groundtruth PSNR, XPSNR, and bitrate to train the prediction models. This example shows the ground truth calculation of a video encoded at 1080p with qp 30. }
\vspace{-0.5em}
\label{fig:gt_xpsnr}
\end{figure}

We used 1000 videos of the Inter-4K dataset~\cite{inter4k_ref} to validate the performance of the encoding methods. We encoded the sequences at UHD (2160p) 60fps using VVenC v1.10~\cite{vvenc_ref} using preset 0 (\textit{faster}).  We extracted the spatiotemporal features, \{\EY, \EU, \EV, \LY, \LU, \LV, \h\} using VCA v2.0~\cite{vca_ref}. We ran constant quality encoding by varying \emph{qp} values from $q_{\text{min}}$ to $q_{\text{max}}$ for each resolution in $\mathcal{R}$. We computed full-reference PSNR and XPSNR quality metrics after the compressed video was upscaled to the original resolution (2160p), as shown in Fig.~\ref{fig:gt_xpsnr}. We used the bi-cubic filtering algorithm for upscaling and downscaling (from 2160p) of videos.

\subsection{Model training}
\paragraph*{XPSNR prediction:} We trained the XPSNR prediction models using multiple regressors, including extra-trees, XGBoost, and random forests, and we observed that the XGBoost regressor~\cite{xgboost_ref} performed the best consistently using our feature set. We performed five-fold cross-validation, and the results are averaged from all folds. Since these values are similar, we assume that the model generalizes well. We ensured the training set did not include segments from the same scenes in the test set. We performed a grid search to explore different combinations of hyperparameter values for the XGBoost model and selected  \emph{max\_depth}=10, and \emph{n\_estimators}=400 that maximized performance.

\paragraph*{QP prediction:} Since machine learning models are associated with an error during prediction, and we need our QP to be monotonic, our QP prediction framework employs a cascading approach to predict bitrate for a given QP.  This method involves training distinct XGboost regression models for minimum and maximum QP values ($q_{\text{min}}$ and $q_{\text{max}}$, respectively), enabling the prediction of the maximum ($\hat{b}_{\text{max}}$) and minimum bitrate ($\hat{b}_{\text{min}}$). Once the points ($q_{\text{min}}$, $\hat{b}_{\text{max}}$)  and ($q_{\text{max}}$, $\hat{b}_{\text{min}}$) are estimated, the optimized QP $\hat{q}$ for a target bitrate $b$ is determined using linear regression~\cite{qadra_ref}, as follows:

\begin{align}
\resizebox{0.90\hsize}{!}{   $\hat{q} = q_{\text{max}} + \Big(\frac{q_{\text{max}} - q_{\text{min}}}{log (\hat{b}_{\text{max}}) - log (\hat{b}_{\text{min}})}\Big) \cdot \Big(log (b) - log (\hat{b}_{\text{max}})\Big)$ }
\end{align}

The equation captures the non-linear relationship between bitrate and QP by employing a logarithmic mapping of the bitrate values. This enables finer granularity in QP adjustment, particularly for target bitrates that fall within the lower or higher ends of the predicted bitrate range. Moreover, considering both the maximum and minimum predicted bitrates ensures stability and reliability in QP prediction, even in scenarios where the target bitrate varies significantly from the predicted bitrate range.

\begin{table}[t]
\caption{Experimental parameters used in this paper.}
\vspace{-0.64em}
\centering
\resizebox{0.905\columnwidth}{!}{
\begin{tabular}{l||c|c|c|c|c|c}
\specialrule{.12em}{.05em}{.05em}
\specialrule{.12em}{.05em}{.05em}
\emph{Parameter} & \multicolumn{6}{c}{\emph{Values}}\\
\specialrule{.12em}{.05em}{.05em}
\specialrule{.12em}{.05em}{.05em}
$\mathcal{R}$ & \multicolumn{6}{c}{\{ 360, 540, 720, 1080, 1440, 2160 \} } \\
\hline
$\mathcal{B}$ & 0.145 & 0.300 & 0.600 & 0.900 & 1.600 & 2.400 \\
              & 3.400 & 4.500 & 5.800 & 8.100 & 11.600 & 16.800 \\
\hline
$r_{\text{max}}$ &  \multicolumn{2}{c|}{720} & \multicolumn{2}{c|}{1080} & \multicolumn{2}{c}{2160} \\
\hline
Target encoder & \multicolumn{6}{c}{VVenC (faster)} \\
\hline
Target decoder & \multicolumn{6}{c}{VTM} \\
\specialrule{.12em}{.05em}{.05em}
\specialrule{.12em}{.05em}{.05em}
\end{tabular}
}
\vspace{-0.98em}
\label{tab:exp_par}
\end{table}

\begin{figure*}[t]
\centering
\begin{subfigure}{0.240\linewidth}
    \centering
    \includegraphics[width=0.98\textwidth]{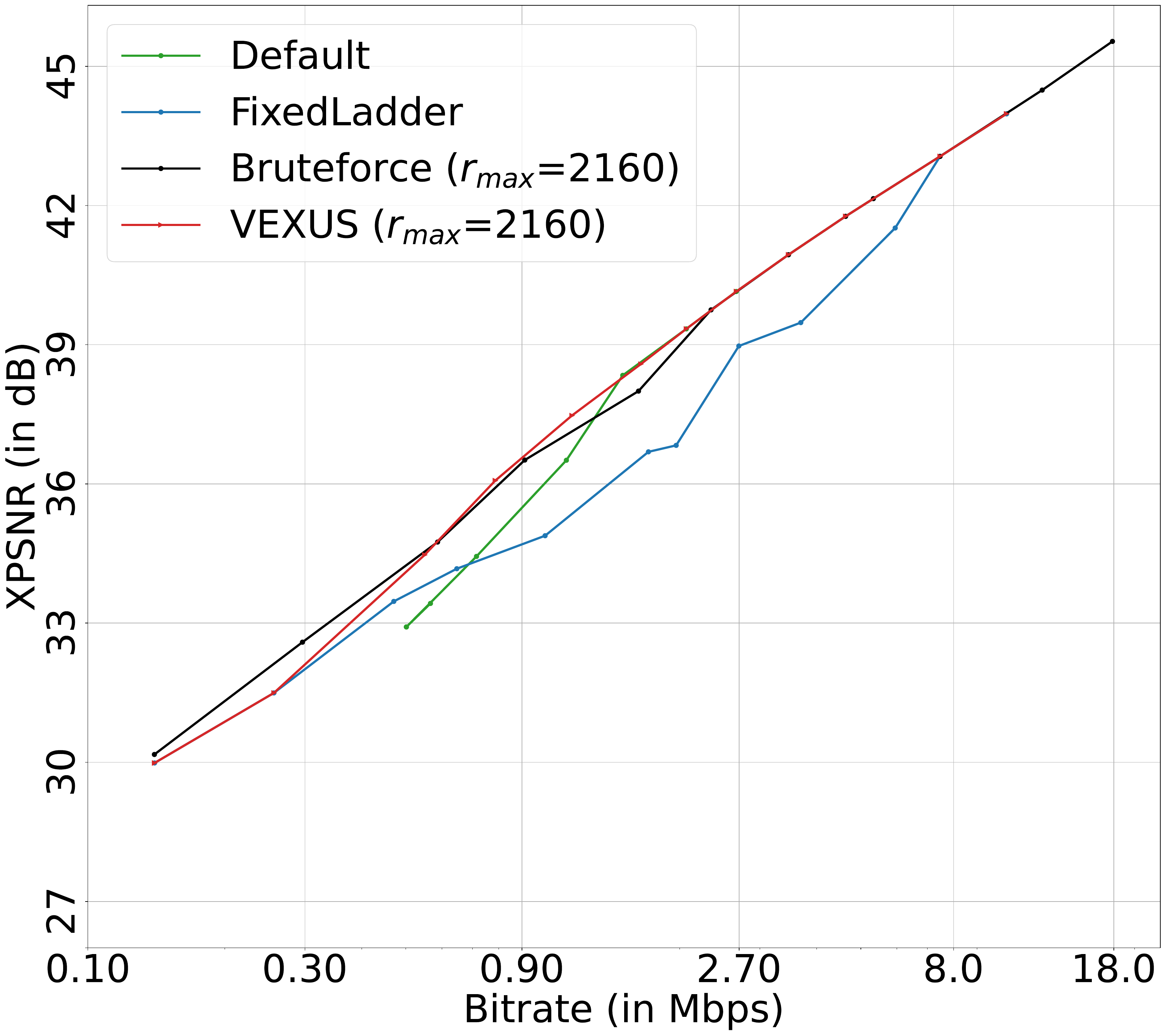}
    \includegraphics[width=0.98\textwidth]{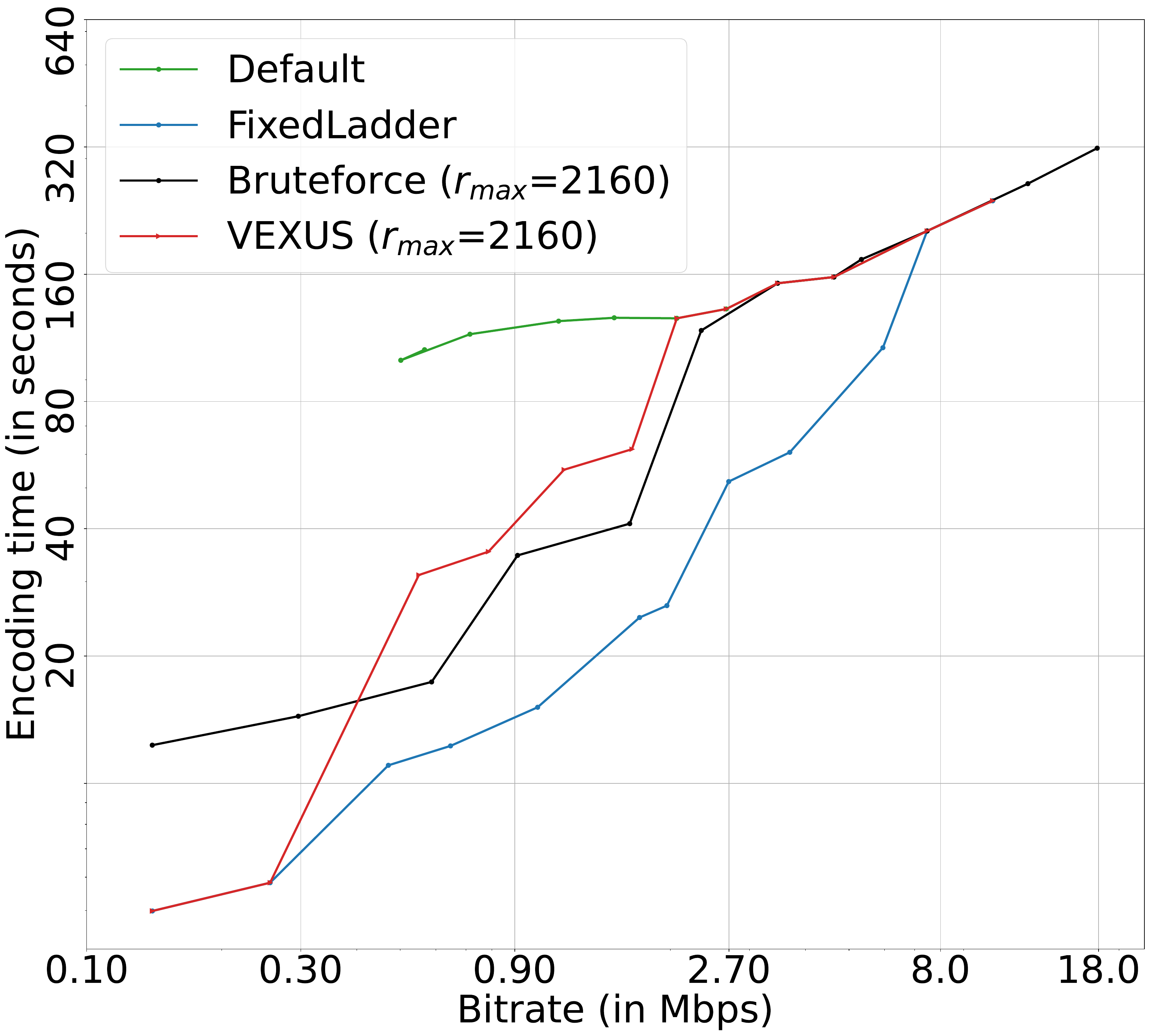} 
    \includegraphics[width=0.98\textwidth]{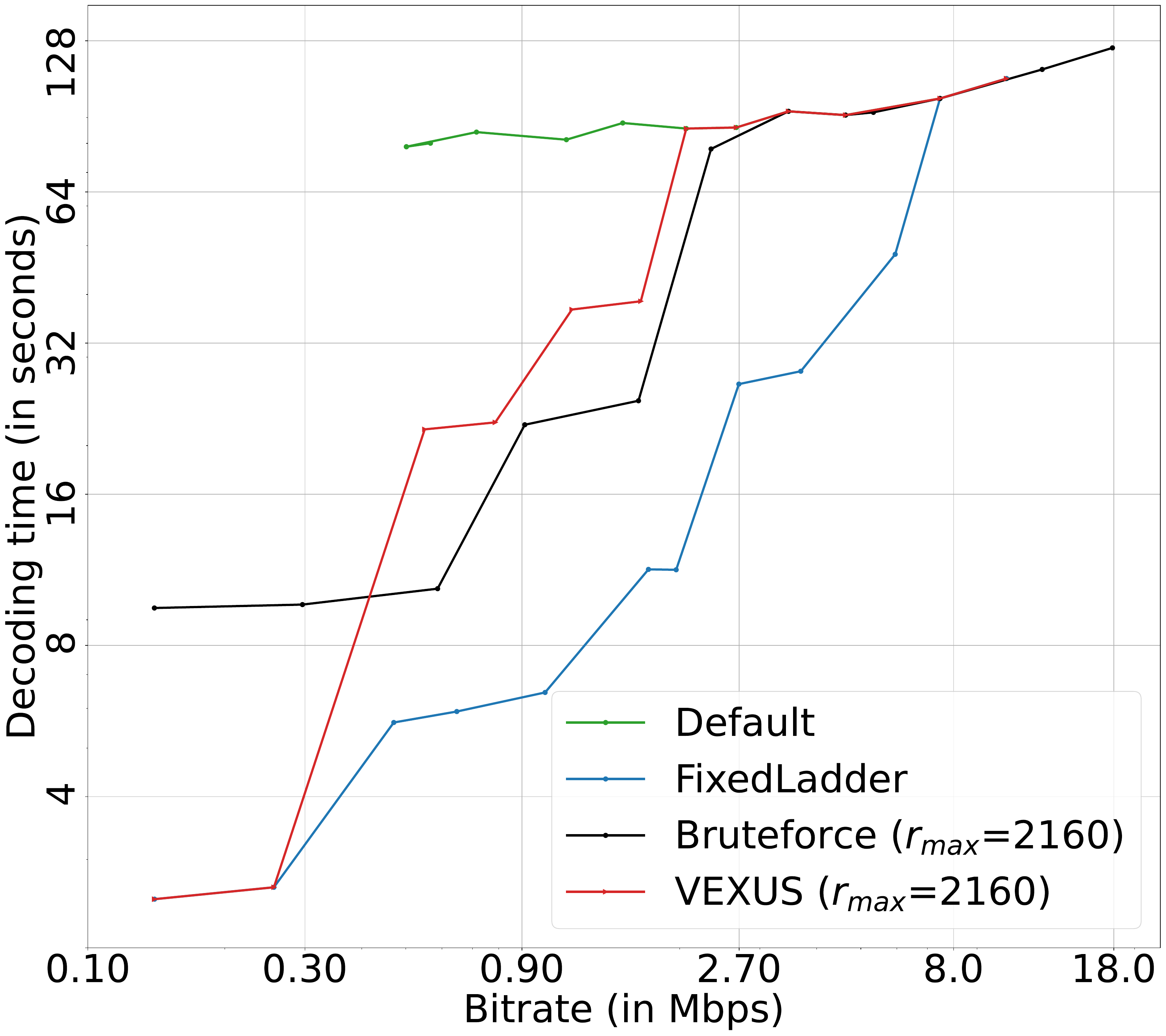}     
    \caption{\textit{0076} (\EY=43.49, \h=21.70, \LY=122.92)}
\end{subfigure}
\begin{subfigure}{0.240\linewidth}
    \centering
    \includegraphics[width=0.98\textwidth]{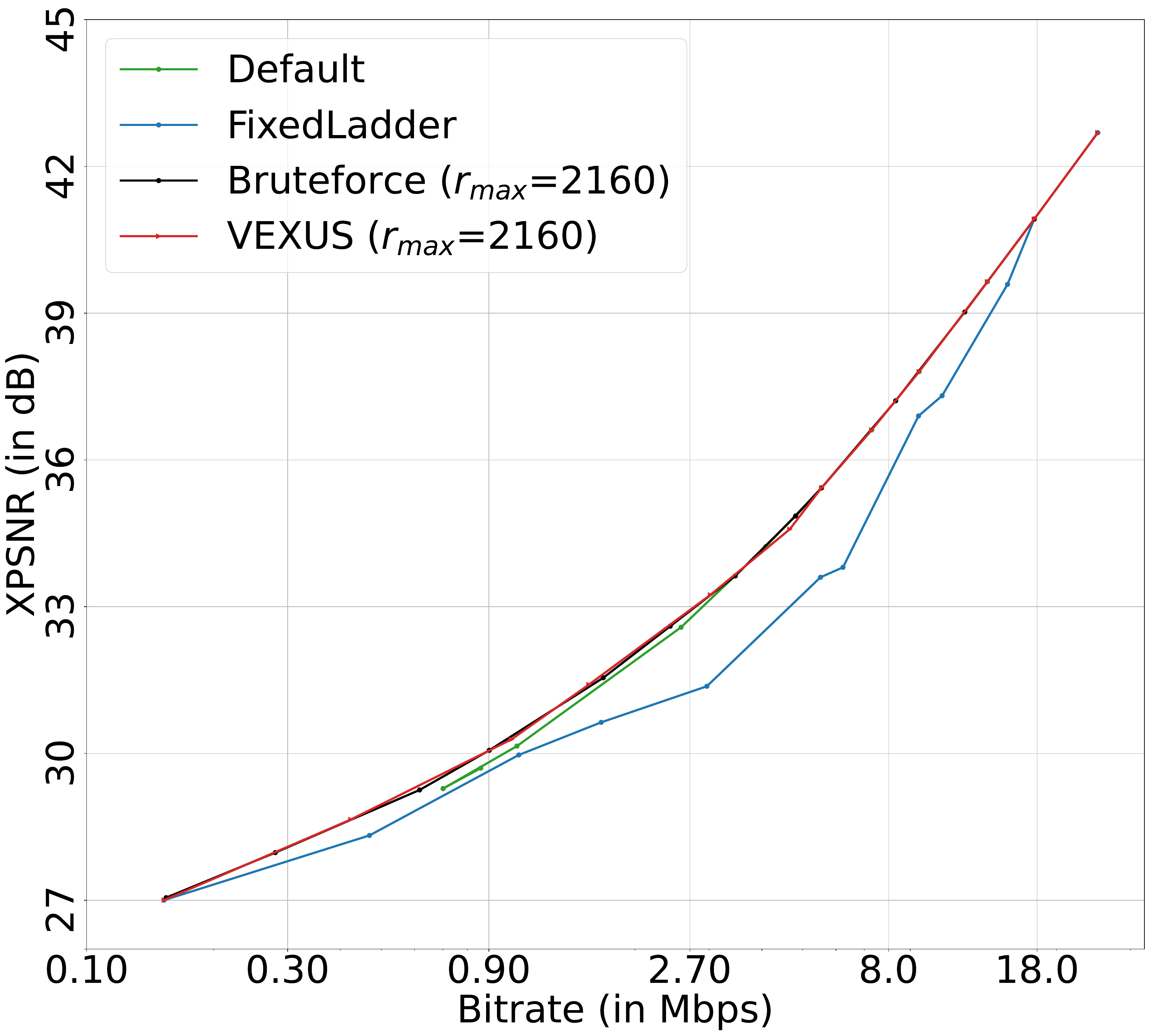}
    \includegraphics[width=0.98\textwidth]{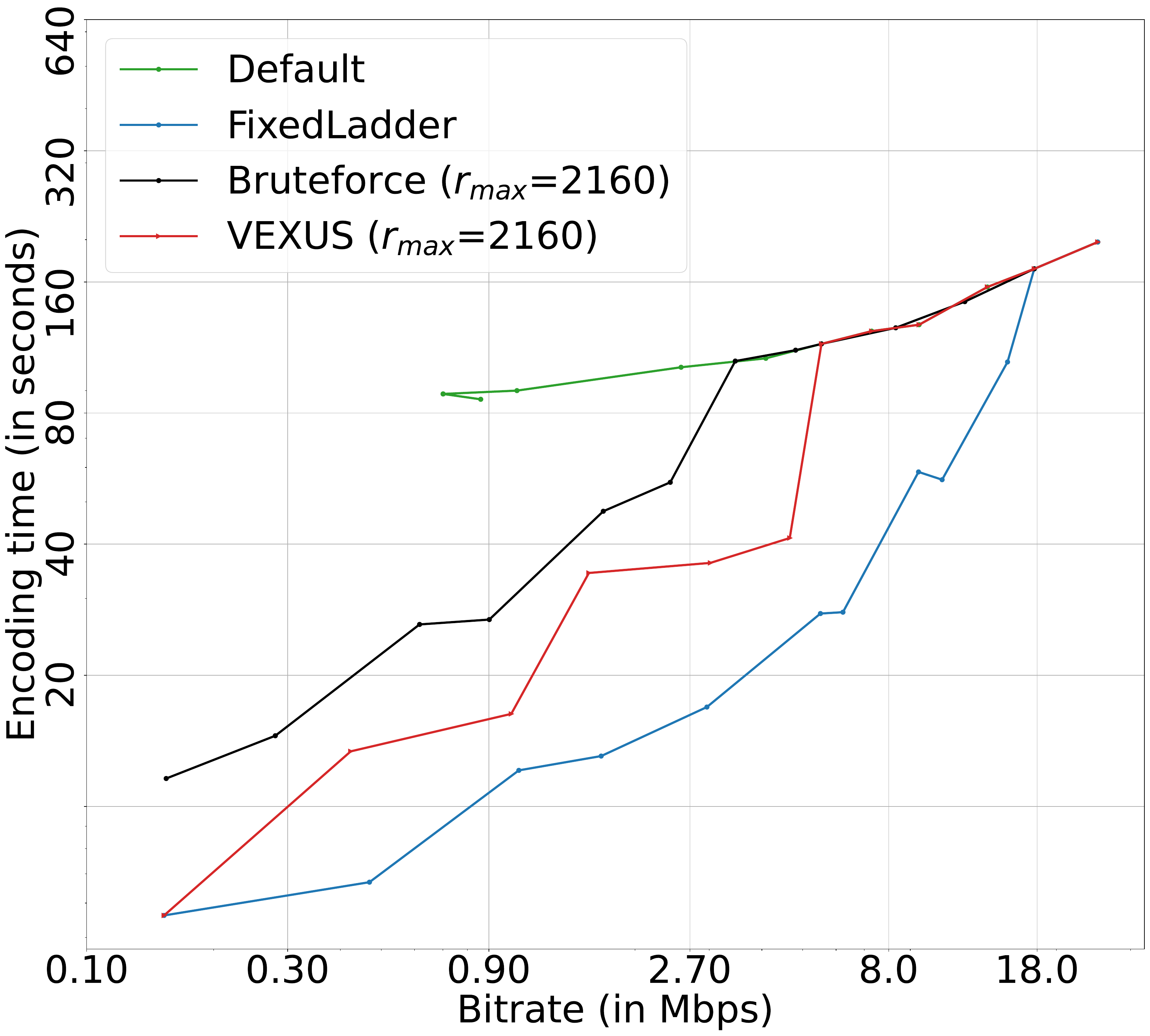}  
    \includegraphics[width=0.98\textwidth]{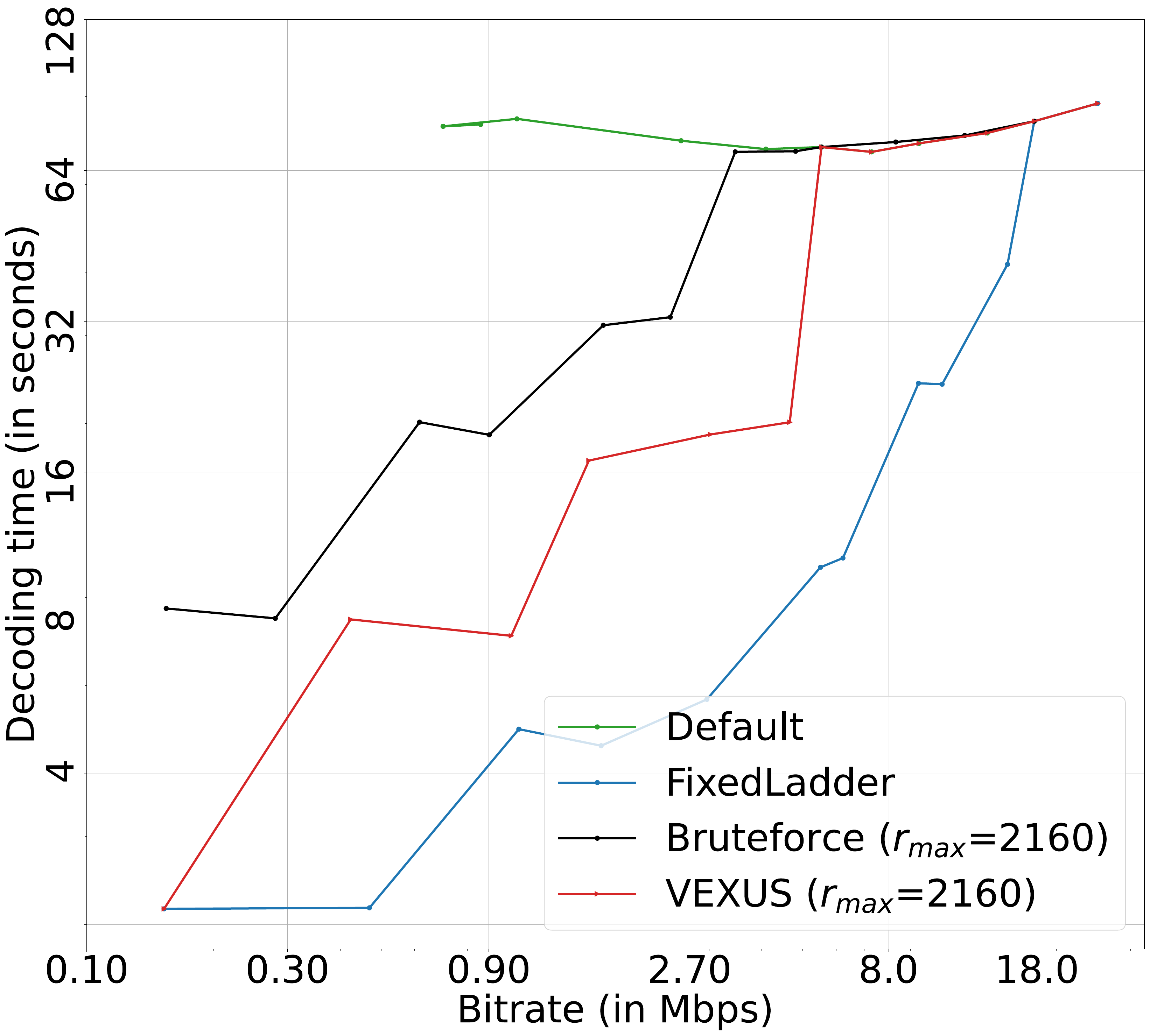}     
    \caption{\textit{0094} (\EY=148.67, \h=14.00, \LY=100.01)
}
\end{subfigure}
\begin{subfigure}{0.240\linewidth}
    \centering
    \includegraphics[width=0.98\textwidth]{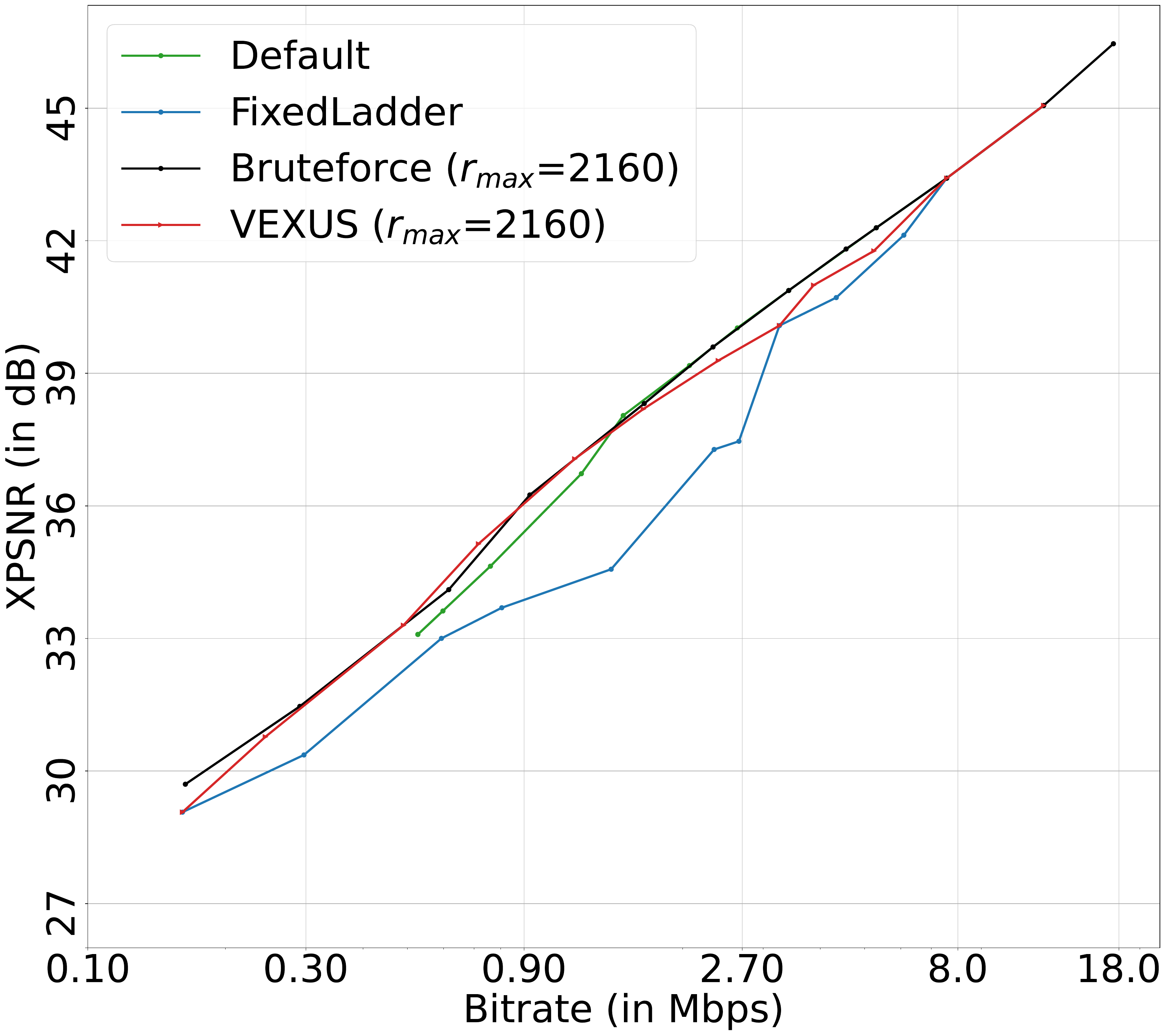}
    \includegraphics[width=0.98\textwidth]{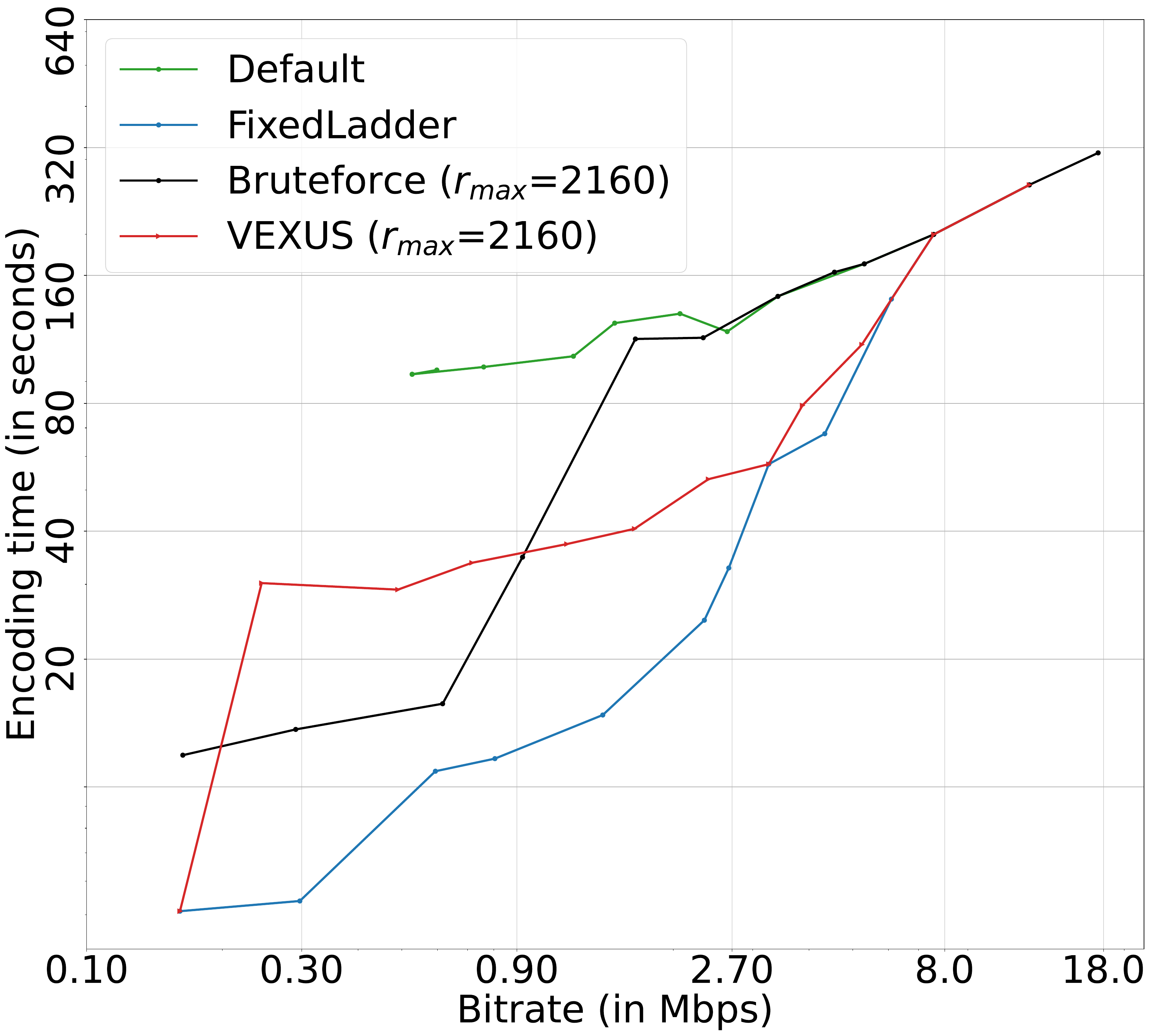}
    \includegraphics[width=0.98\textwidth]{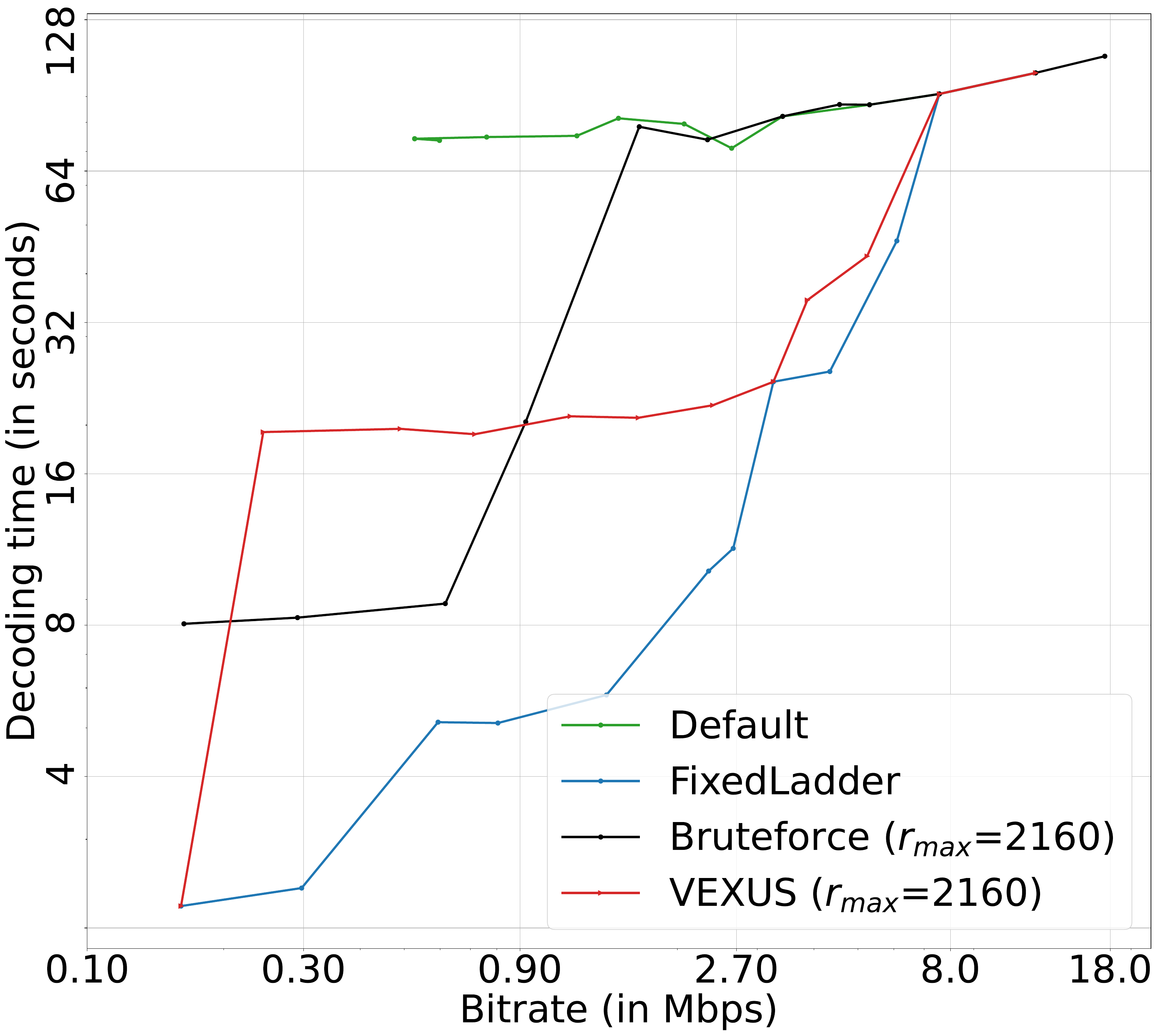}    
    \caption{\textit{0153} (\EY=50.83, \h=9.80, \LY=93.54)}
\end{subfigure}
\begin{subfigure}{0.240\linewidth}
    \centering
   \includegraphics[width=0.98\textwidth]{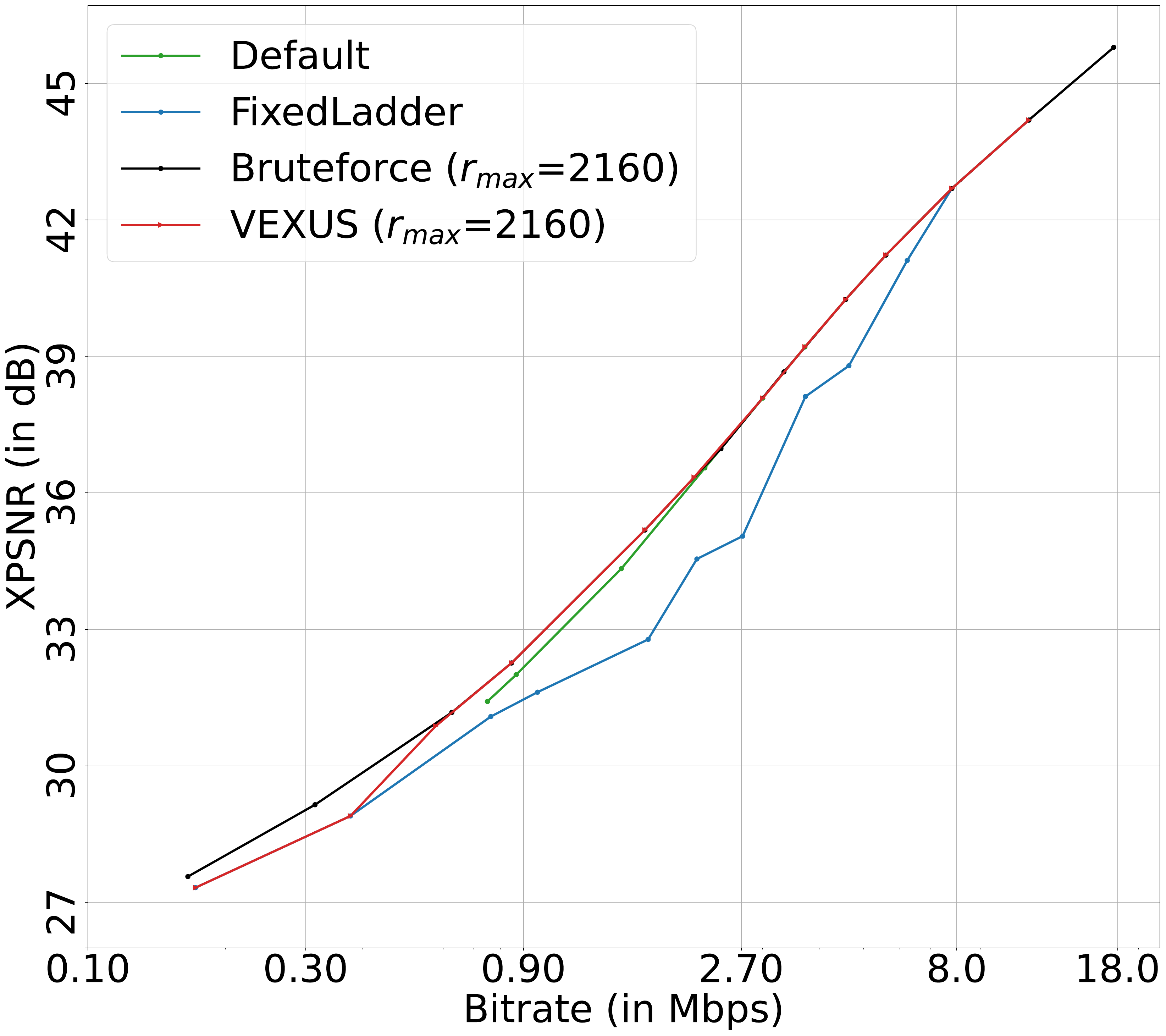} 
   \includegraphics[width=0.98\textwidth]{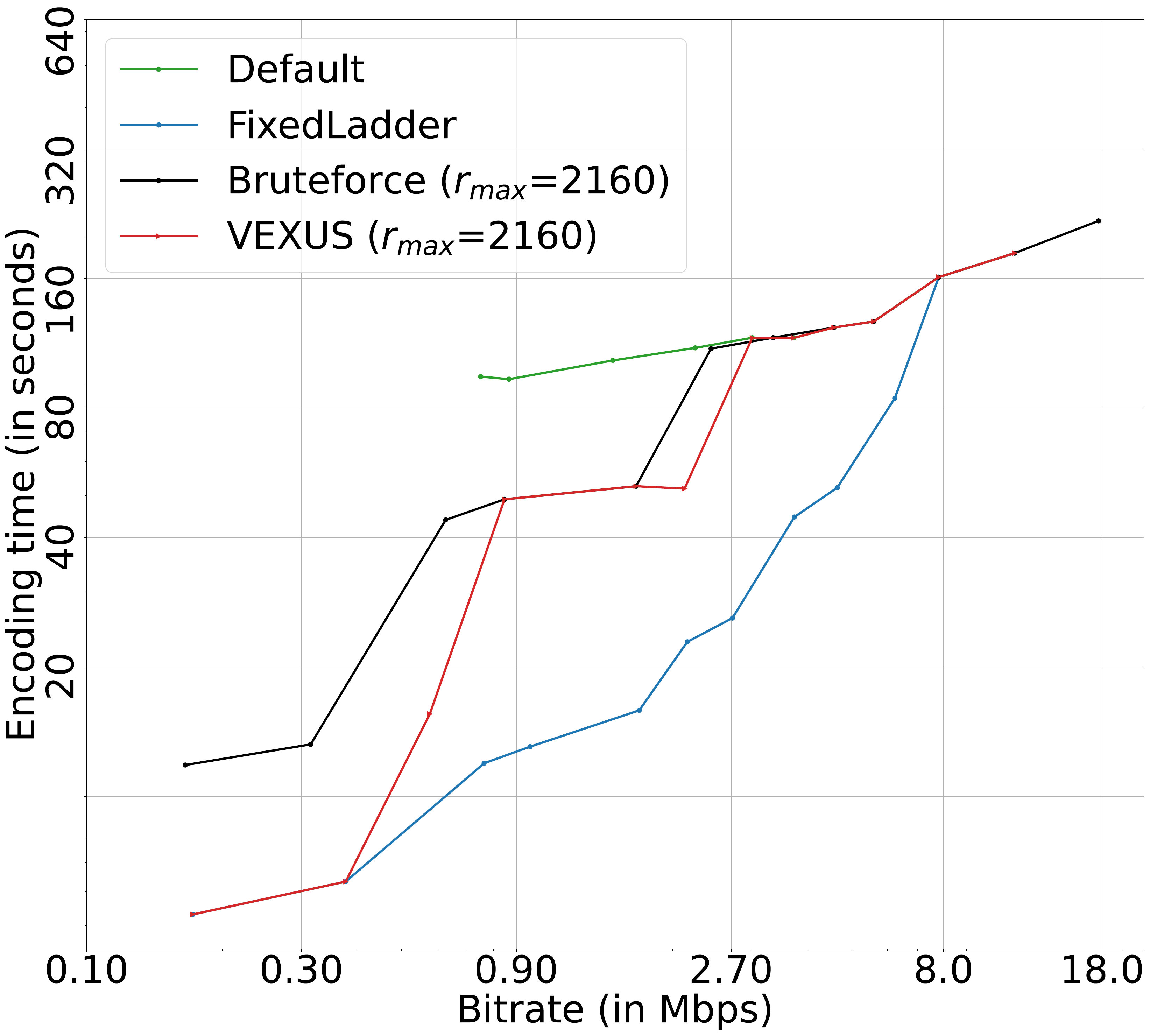}   
   \includegraphics[width=0.98\textwidth]{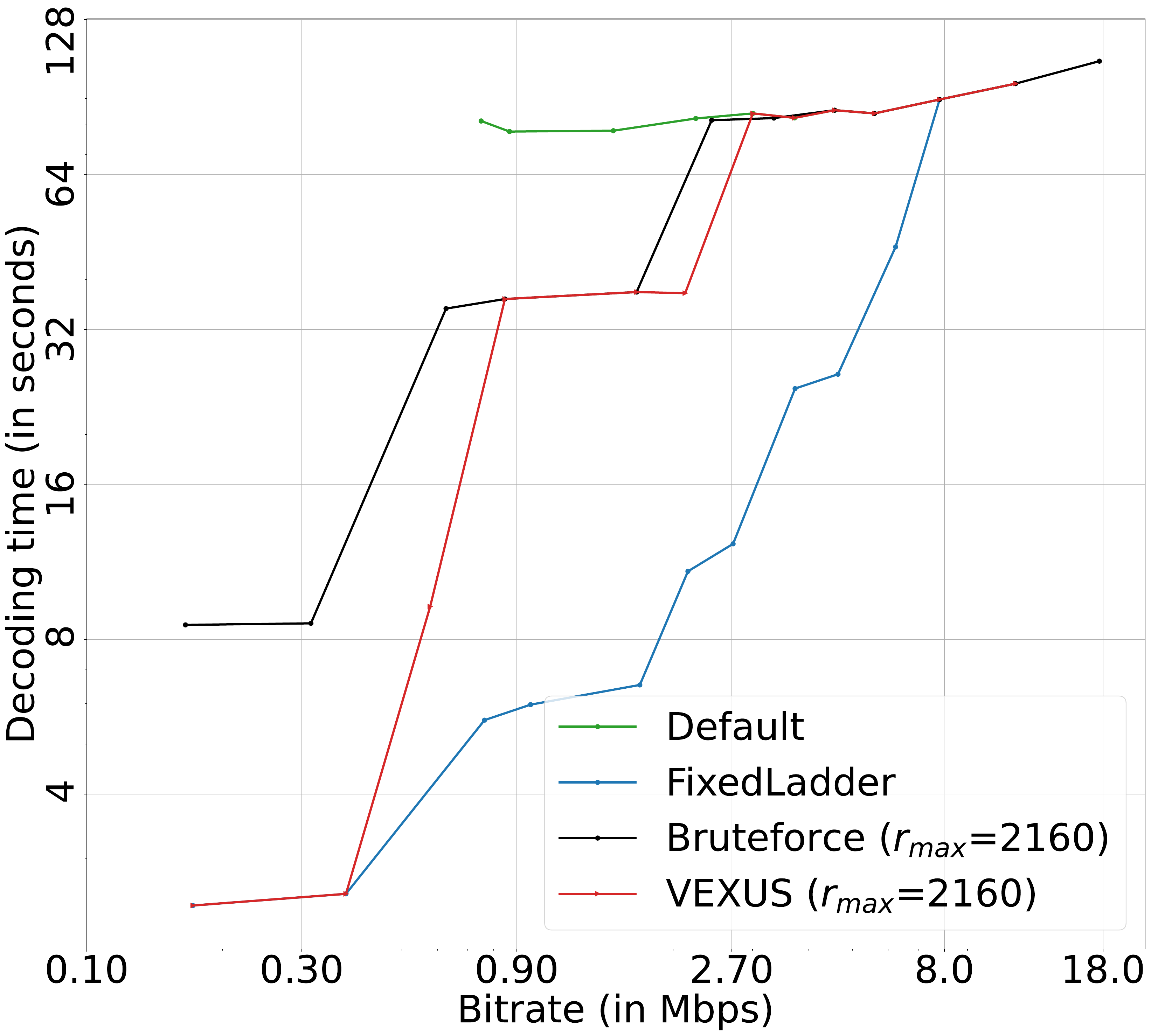}     
    \caption{\textit{0263} (\EY=122.64, \h=38.48, \LY=116.31)}
\end{subfigure}
\vspace{-0.6em}
\caption{RD curves, encoding and decoding times of representative video sequences using \textit{Default} (green line), \textit{FixedLadder} (blue line), \textit{Bruteforce} (black line), and \scheme (red line) encodings.}
\vspace{-0.8em}
\label{fig:rd_res}
\end{figure*}

\subsection{Evaluation setup} 
We run all experiments on a dual-processor server with Intel Xeon Gold 5218R (80 cores, frequency at 2.10 GHz), where each VVenC encoding instance uses four CPU threads. We use $q_{\text{min}}$ and $q_{\text{max}}$ as 10 and 50, respectively. We run VCA using four CPU threads. The experimental parameters used to evaluate \scheme are shown in Table~\ref{tab:exp_par}.

\paragraph*{Reference methods: } 
We compare \scheme with the following benchmarks:
\begin{enumerate}
    \item \emph{Default}: This method employs a fixed resolution encoding, \ie all bitstreams are encoded at the exact resolution as the input video.
    \item \emph{FixedLadder}: This method employs a fixed set of bitrate-resolution pairs. This paper uses the HLS bitrate ladder specified in the Apple authoring specifications~\cite{HLS_ladder_ref} as the fixed set of bitrate-resolution pairs.
    \item \emph{Bruteforce}: This method determines optimized resolution, which yields the highest \mbox{XPSNR} for a given target bitrate after an exhaustive encoding process at all supported resolutions and QPs~\cite{netflix_paper,chen_pte_ref}.
\end{enumerate}

\paragraph*{Performance metrics: } 
We compare the overall quality in PSNR~\cite{psnr_ref} and XPSNR~\cite{xpsnr_ref} and the achieved bitrate for every target bitrate of each test sequence. Bjøntegaard Delta values~\cite{DCC_BJDelta} \mbox{BD-PSNR} and \mbox{BD-XPSNR} refer to the average increase in \mbox{PSNR} and \mbox{XPSNR} of the bitstreams compared with the \emph{Default} encoding with the same bitrate, respectively. Bjøntegaard Delta rate \BDRP~and \BDRX~refer to the average increase in bitrate of the bitstreams compared with the \emph{Default} encoding with the same \mbox{PSNR} and \mbox{XPSNR}, respectively.  We also evaluate the average relative difference in encoding time ($\Delta T_{\text{E}}$) and decoding time ($\Delta T_{\text{D}}$) compared with the \emph{Default} encoding.


\section{Evaluation results}
\label{sec:results}
\paragraph*{Speed and accuracy:} 
We evaluate the pre-processing latency ($\tau_{\text{p}}$) in encoding introduced by the spatiotemporal complexity feature extraction and the model inference to predict the optimized resolutions and QPs. We extract the features at an average rate of 178\,fps over the entire dataset (2160p resolution). The time to predict the resolution for each target bitrate is \SI{4}{\milli\second}. The overall latency is negligible as spatiotemporal complexity feature extraction, and the XPSNR-optimized resolution prediction can be executed concurrently in real applications. The average mean absolute error (MAE) of the XPSNR prediction model is \SI{0.17}{\decibel} with a standard deviation of \SI{0.22}{\decibel}, while the MAE of the QP prediction model is 1.32, with a standard deviation of 1.86.

\paragraph*{Rate-distortion performance:} 
Fig.~\ref{fig:rd_res} shows the RD curves of the representative videos (of various video content complexities) in the test dataset. Notably, the \textit{Bruteforce} method consistently yields the highest XPSNR, showcasing its exhaustive search approach. Meanwhile, the RD curve of \scheme closely mirrors the \textit{Bruteforce} method, indicating the effectiveness of its predictive modeling in approximating optimized resolutions and QPs. Furthermore, Table~\ref{tab:res_cons} shows \BDRP, \BDRX, \mbox{BD-PSNR}, and \mbox{BD-XPSNR} results of the considered methods compared to the \emph{Default} encoding. \scheme ($r_{\text{max}}$=2160) yields \BDRP, \BDRX, \mbox{BD-PSNR}, and \mbox{BD-XPSNR} of \SI{-25.66}{\percent}, \SI{-18.18}{\percent}, \SI{5.84}{\decibel}, and \SI{0.62}{\decibel}. It is observed that the coding efficiency decreases as $r_{\text{max}}$ decreases. \vig{To further validate the efficiency of \scheme, we evaluated the perceptual quality using the VMAF metric in addition to PSNR and XPSNR. \mbox{BD-VMAF} of \emph{FixedLadder}, and \scheme ($r_{\text{max}}$=2160) are  \SI{-2.17}{} and \SI{9.39}{}, which confirms the robustness of \scheme.}

\begin{table}[t]
\caption{Average results of the encoding schemes compared to \textit{Default} encoding.}
\vspace{-0.5em}
\centering
\resizebox{1.00\linewidth}{!}{
\begin{tabular}{l|c||c|c|c|c|c|c}
\specialrule{.12em}{.05em}{.05em}
\specialrule{.12em}{.05em}{.05em}
Method & $r_{\text{max}}$ & \makecell{\BDRP} & \BDRX &  BD-PSNR & BD-XPSNR & $\Delta T_{\text{E}}$ & $\Delta T_{\text{D}}$  \\
& [pixels] & [\%] & [\%] &  [dB] & [dB] & [\%] & [\%] \\
\specialrule{.12em}{.05em}{.05em}
\specialrule{.12em}{.05em}{.05em}
\textit{FixedLadder} & - & -11.86 & 1.35 & 1.53 & -0.24 & -52.39 & -65.77\\
\hline
\multirow{3}{*}{\textit{Bruteforce}} & 720 & -13.17 & -6.29 & 4.65 & 0.51 & 418.96 & -78.82 \\
 & 1080 & -19.49 & -7.93 & 6.21 & 0.83 & 596.23 & -72.60 \\
 &  2160 & -27.26 & -20.27 & 7.45 & 1.02 & 1312.37 & -65.12 \\
\hline
\multirow{3}{*}{\scheme} & 720 & -11.24 & -3.17 & 2.93 & 0.22 & -73.97 & -79.06 \\
 &  1080 & -17.89  & -6.01  & 4.30  & 0.55 & -64.71 & -74.27 \\
 &  2160 & -25.66  & -18.80 &  5.84 & 0.62 & -44.43 & -65.46 \\
\specialrule{.12em}{.05em}{.05em}
\specialrule{.12em}{.05em}{.05em}
\end{tabular}
}
\vspace{-0.73em}
\label{tab:res_cons}
\end{table}

\paragraph*{Latency:}
As resolutions decrease for a given target bitrate in \scheme, the encoding time tends to decrease. This is because lower resolutions require less computational effort for encoding than UHD resolution. Similarly, decoding time is reduced for lower resolutions as decoding operations become less complex. Therefore, Fig.~\ref{fig:rd_res} demonstrates that selecting lower resolutions for a given target bitrate not only improves encoding efficiency but also lowers both encoding and decoding time compared to the \emph{Default} encoding, making it a favorable approach for optimizing VVC encoding. As shown in Table~\ref{tab:res_cons}, \textit{Bruteforce} approach yields the highest encoding time, as we need to encode at all supported resolutions and QPs to determine the convex-hull. In \scheme, the highest encoding time is observed when $r_{\text{max}}$=2160 due to selecting higher encoding resolutions optimized for maximum XPSNR. As $r_{\text{max}}$ decreases, the overall encoding and decoding time decreases, and the coding efficiency is lowered. This trade-off between quality and coding efficiency is based on the target audience, delivery platform, and available resources. 

\section{Conclusions}
This paper introduced XPSNR as a\vig{n alternative} metric to the traditional VMAF metric to \vig{estimate} convex-hull for VVC encoding. Our investigation stemmed from a keen observation: XPSNR demonstrated a better correlation with subjective quality scores for VVC-coded UHD content. Leveraging this insight, we introduced an approach where XPSNR is predicted for VVC-coded bitstreams using spatiotemporal complexity features of the video and the target encoding configuration. Furthermore, we propose \scheme, where the convex-hull is estimated online using the predicted XPSNR. On average, \scheme yields a substantial improvement of \SI{5.84}{\decibel} in PSNR and \SI{0.62}{\decibel} in XPSNR for the same bitrates compared to the conventional UHD encoding with the VVenC encoder, followed by a \SI{44.43}{\percent} reduction in overall encoding time, and a \SI{65.46}{\percent} reduction in overall decoding time using VTM decoder. As the demand for superior streaming experiences grows, our findings emphasize the pivotal role of XPSNR in achieving optimized bitrate-resolution pairs for VVC-coded content.

One promising avenue of future research is investigating how the insights of convex-hull can be leveraged within the encoder's rate control mechanisms and mode decisions. By incorporating knowledge of the convex hull estimation into rate control algorithms, the encoder could dynamically adjust encoding parameters to better allocate bitrate resources, particularly in scenarios where low resolutions are utilized to enhance quality at lower bitrates.
\balance
\bibliographystyle{IEEEtran}
\bibliography{references.bib}
\balance
\end{document}